# A Scalable and Robust Framework for Data Stream Ingestion

Haruna Isah
School of Computing
Queen's University
Kingston, Canada
h.isah@cs.queensu.ca

Farhana Zulkernine
School of Computing
Queen's University
Kingston, Canada
farhana@cs.queensu.ca

*Abstract*—An essential part of building a data-driven organization is the ability to handle and process continuous streams of data to discover actionable insights. The explosive growth of interconnected devices and the social Web has led to a large volume of data being generated on a continuous basis. Streaming data sources such as stock quotes, credit card transactions, trending news, traffic conditions, time-sensitive patient's data are not only very common but can rapidly depreciate if not processed quickly. The ever-increasing volume and highly irregular nature of data rates pose new challenges to data stream processing systems. One such challenging but important task is how to accurately ingest and integrate data streams from various sources and locations into an analytics platform. These challenges demand new strategies and systems that can offer the desired degree of scalability and robustness in handling failures. This paper investigates the fundamental requirements and the state of the art of existing data stream ingestion systems, propose a scalable and fault-tolerant data stream ingestion and integration framework that can serve as a reusable component across many feeds of structured and unstructured input data in a given platform, and demonstrate the utility of the framework in a real-world data stream processing case study that integrates Apache NiFi and Kafka for processing high velocity news articles from across the globe. The study also identifies best practices and gaps for future research in developing large-scale data stream processing infrastructure.

I. INTRODUCTION

With the evolution and increasing popularity of social media platforms and Internet of Things (IoT), inconceivable volumes of data are being generated [1]. Internet Protocol (IP) traffic data as predicted by Ballard *et al* [2] have now reached half a zettabyte. Greater portion of this data may not have an apparent use today but may be useful in the future. Some of this data may lose its value or be lost forever if not processed immediately. It is, therefore, important to ingest, process, save the important aspects of the data and delete the portions that are not useful [3]. With the way new data sources are evolving daily, more businesses will depend on being able to process and make decisions on data streams.

The challenge of data explosion has generated a lot of interest in low-latency in-memory frameworks and data stream processing systems in recent years [4]. In order to keep up with the high rate and volume of data, modern data processing systems must deliver insights with minimal latency and high throughput [5]. Streaming data analytics is a new programming paradigm designed to incorporate continuous data into a decision-making process [2]. It is useful in identifying perishable insights which require immediate or time constraint action. However, as a variety of data stream processing tools have become available, understanding the required capabilities of streaming architectures is vital to making the right design or usage choices [6].

A typical streaming analytics system is built on top of a three layers stack that include ingestion, processing, and storage components [5]. The ingestion layer is the entry point to the streaming architecture. It decouples, automates, and manages the flow of information from data sources to the processing and storage layers. The processing layer consumes the data streams buffered by the ingestion layer and sends the output or intermediate results to the storage layer. The storage layer is responsible for holding data in an in-memory data store for iterative computations or in databases for long-term persistence [6, 7]. The stored data may be processed further and the analytics results are delivered to a variety of display and decision support tools [8].

Data ingestion is an area that is often overlooked, yet its importance cannot be underestimated [1]. Many organizations have data stored in files that need to be moved around and processed at many different locations. Data in motion needs dataflow management [9]. Traditionally streaming analytics systems are mostly limited to handling dataflows within a local data center. However, the world has become more connected to the extent that many organizations now operate over several data centers in different geo-locations. Streaming analytics systems are faced with the challenge of collecting and connecting huge data streams across the globe. This is an important requirement in big data projects where companies aim to ingest a variety of data sources ranging from live multimedia, to IoT data, and to real-time headlines from social media and blogs. Another challenge is to provide security, auditing, and provenance in a data ingestion mechanism. The analytical value of data entirely depends on its completeness, accuracy, and consistency. Achieving accurate and continuous data ingestion and management is a complex and challenging task that requires proper planning, specialized tools, and expertise [10]. These challenges demand new strategies and systems that can offer the desired degree of scalability and robustness in handling failures. Data ingestion research has previously been targeted under different research initiatives such as Extract, Transform, and Load (ETL), data integration, deduplication, integrity constraint maintenance, and bulk data loading [11]. Data flow management is used in this study to refer to the tasks of ingesting, integrating, extracting, enriching, and distributing data streams within or outside an analytic platform.

The contributions of this work are as follows. First, we propose a scalable and fault-tolerant dataflow management framework that can serve as a reusable component across many feeds of structured and unstructured input data. Second, we demonstrate the utility of the framework in a real-world data stream processing case study that integrates Kafka and HDFS in a dataflow system powered by NiFi. The paper is organized as follows. Section II introduces the requirements of a data stream ingestion system. Following the requirements, we present the proposed framework and its features in section III. Section IV details our experimental and evaluation results. A study of related work is presented in section V. Finally, we give concluding remarks in Section VI.

## II. REQUIREMENTS OF A STREAM INGESTION SYSTEM

Data ingestion layer serves to acquire, buffer and optionally pre-process data streams (e.g., filter) before they are consumed by the analytics application. Important features when considering data stream ingestion tools include the ease of installing, publishing, transporting, consuming, and archiving streams to disk. Data ingestion system should be able to support high throughput, low latency and must scale to a large number of both data stream producers and consumers [12]. We categorized these requirements to include source integration and pre-processing, fault tolerance and message delivery guarantees, provenance and security, backpressure and routing, scalability, and extensibility.

### A. Source integration and preprocessing

Data streams may be sourced directly from http/Web Sockets Application Programming Interface (API), REST API, Streaming API, IoT Hubs, or through message queuing sources. One of the fundamental issues of stream computing is the challenge of collecting and integrating data from a multitude of sources. The most complex and difficult part of integrating data from various sources is the task of transforming data into a common format [13]. When the source of streaming data is diverse, for instance hundreds of sources emitting dozens of data formats, improving the rate of data ingestion and efficiency of data processing becomes a challenging task [10]. Multiple applications may also be lined up to consume the ingested data. It is desirable to integrate the incoming streams into a single flow and then transform it in multiple ways to drive different applications concurrently [7].

Effective data stream ingestion involves the prioritization of data sources, the validation of individual files and the routing of data items to a desired destination. Data stream ingestion systems should be able to verify and filter sources, language, and content format to ensure that the source integration is accurate, smooth, and free from noise (such as duplicates). Some pre-processing steps will be required to integrate global news data streams from various sources such as RSS feeds, blogs, and social media which are mainly unstructured in nature. The technology for handling real-time data integration is more complex than that of static data [13].

### B. Fault tolerance and message delivery guarantees

Data ingestion is expected to run on a large cluster that may be prone to software, hardware, and other third-party system failures. Failures can cause the loss of large amounts of data streams which may lead to erroneous analytic results. Ingestion systems should incorporate high-availability mechanisms that allow to operate continuously for a long time in spite of failures [5]. It is, therefore, desirable for ingestion systems to offer the desired degree of robustness in handling failures while minimizing data loss [7]. Data ingestion needs to support high throughput, low latency and must scale to a large number of both data producers and consumers [12].

### C. Provenance and security

One of the reasons modern systems can now more easily handle streaming data is improvements in the way message-passing systems work. Highly effective messaging technologies collect streaming data from many sources—sometimes hundreds, thousands, or even millions—and deliver it to multiple consumers of the data, including but not limited to real-time applications. Effective message-passing capabilities are needed as a fundamental aspect of the streaming infrastructure.

### D. Scalability

Multiple feeds with variable data arrival rates imply a varying demand for resources. An ingestion system should be scalable to be able to ingest increasingly large volumes of data from multiple sources. The system should also demonstrate elasticity through an automatic scaling in or out in order to meet the varying demand for resources [7].

### E. Backpressure and Routing

Backpressure in data stream processing is a situation where an ingestion or other components of a DSPS is unable to handle the rate at which the data streams are received. An ingestion system should be able to buffer data in the case of temporary spikes in workload and provide a mechanism to replay it later. Rate throttling is a typical example of backpressure mechanism for handling fast arrivals of data streams. It is an artificial restriction to the rate at which tuples are delivered. [5].

### F. Extensibility

Ingestion systems must be generic enough to work with a variety of data sources and high-level applications. A plug-and-play model is desired to allow extension of the offered functionality [7]. This provides the ability to add or remove consumers or new functionalities at any time without changing the data ingestion pipeline. Ingestion systems should be able to filter any erroneous or malicious data items before transporting the data to processing or storage layers.

## III. PROPOSED FRAMEWORK

The focus of this paper is dataflow management which refers to an automated and managed flow of information between systems. Our goal is to design a highly-flexible dataflow management framework for large-scale data stream ingestion and integration with the capabilities of meeting all or most of the challenges detailed in the previous section. The proposed dataflow management framework is shown in Fig. 1. The framework components (in blue) include (1) Data stream acquisition from disparate sources (2) Data stream extraction, enrichment, and integration and (3) Data stream distribution to various downstream systems such as data store or analytics platform.

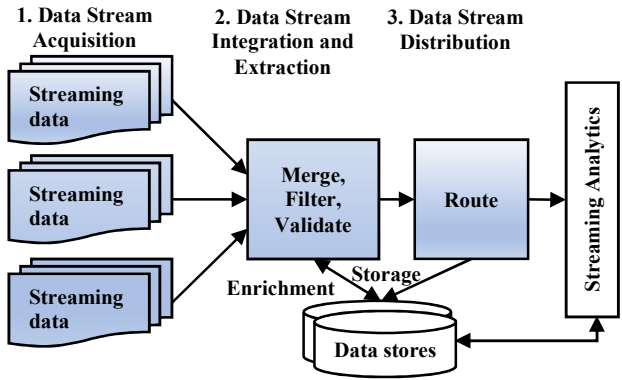

Figure 1. Dataflow management framework

The following subsections describe each of these functionalities in the framework.

*A. Data Stream Acquisition*

The data acquisition is the entry point for bringing data into a processing platform. It involves the ingestion of data of different formats and from several sources. Streams of data such as feeds from IoT devices or Twitter firehose arrive into the system over sockets from internal or external servers. These streams are then merged, filtered and distributed to connected layers for processing, temporary storage, and persistence. The framework supports commonly used data access schemes such as sockets, Representational State Transfer (REST) API, Streaming API, or custom schemes and modern devices or application interaction patterns such as publish-subscribe and stream protocols.

Rather than designing a new tool for data acquisition from the scratch, we are using an open source dataflow management system called NiFi[1]. Young et al [14] describe NiFi as a data in motion technology that uses flow-based processing. It enables data acquisition, basic event processing, and data distribution mechanism. NiFi gives organizations a distributed and resilient platform for building enterprise dataflows [15]. It provides the capability to accommodate diverse dataflows being generated by the connected world. NiFi enables seamless connections among databases, big data clusters, message queues, and devices. It incorporates visual command, control, provenance (data lineage), prioritization, buffering (back pressure), latency, throughput, security, scalability, and extensibility mechanisms [9]. We chose NiFi because it is highly configurable and provides a scalable and robust solution to handle the flow and integration of data streams of different formats from different sources through a cluster of machines. NiFi was designed to meet dataflow challenges which include network failures and crashes, excess loads, corrupt data, rapid changes in organizational requirements, compliance, and security. MiNiFi [2] is an interesting project aimed at extending NiFi's capabilities by collecting data at the edge or source of its creation and bringing it directly to a central NiFi instance.

For the purposes of our study, NiFi will enable us to quickly build simple pipelines for prototyping, before scaling to full production. The above and many features of NiFi, therefore, meets many data acquisition use case requirements.

*B. Data Stream Extraction, Enrichment, and Integration*

Depending on the nature of the incoming data and intended applications, several tasks such as language, noise, and duplicate detection, content parsing, data type transformations are performed on the ingested stream.

*1) Extraction*

Data streams can be ingested in its raw format onto different schemas to enable a variety of different kinds of downstream analytics. The NiFi Expression Language provides the ability to reference attributes, compare them to other values, and manipulate their values. It supports several data types including number, decimal, date, Boolean, and string. It is used heavily throughout the NiFi application for configuring processor properties and provides many different functions (that takes zero or more arguments) to meet the needs of an automated dataflow. The functions can be chained together to create expressions for effective and efficient data stream extraction and manipulations. Near-duplicate detection of incoming data streams is a fundamental extraction task for effective stream ingestion. NiFi provides customizable processors such as DetectDuplicate, for detecting multiple copies of same record in a dataflow; ExecuteScript and ExecuteStreamCommand, for deduplication and filtering of erroneous/malicious data items before transporting the data to processing or storage layers.

*2) Enrichment*

Enrichment is a common use case when working on data ingestion or flow management. It involves getting data from external source (such as database, file, or API) to add more details, context or information to data being ingested. Often the enrichment is done in batch using the join operation. However, doing the enrichment on data streams in real-time is more interesting. NiFi provides processors such as ISPEnrichIP, LookupAttribute, and LookupRecord for data stream enrichment tasks.

*3) Integration*

The practice associated with managing data that travels between applications, data stores, systems, and organizations is traditionally called data integration [13]. Several techniques for managing and integrating data in motion have been developed to decrease the complexity of interactions and increase scalability [6]. Depending on the nature of the incoming data stream, integration may be achieved at the data acquisition stage. Data integration within NiFi is achieved using processors such as MergeContent, MergeRecord, and PartitionRecord.

---

[1] https://nifi.apache.org/

[2] https://nifi.apache.org/minifi/

*C. Data Stream Distribution*

Big data ingestion is about moving data (especially unstructured data) from its sources or producers into big data stores (such as HDFS or Cassandra) or a processing system (such as Storm, Flink, or Spark Streaming) via message queuing systems such as (MQTT and Kafka). Although, there are custom processors for connecting NiFi to big data stream processing engines such as Flink and Spark Streaming. However, using NiFi for delivering massive and high velocity data streams to these systems in a complex multilevel analytics pipeline is not a good practice. This is because if we were to introduce a new consumer of data, for example, a Spark-streaming job, the flow must be changed [1].

We choose Kafka, a high-throughput distributed messaging system which has recently become one of the most common landing places for data within an organization. A common scenario is for NiFi to act as a Kafka producer. In this case, MiNiFi can be utilized to bring data from sources directly to a central NiFi instance, which can then deliver data to the appropriate Kafka topic. A more complex but interesting scenario is a bi-directional flow which combines the power of NiFi, Kafka, and a stream processing engine to create a dynamic self-adjusting dataflow. NiFi-Kafka integration provides the ability to add and remove consumers at any time without changing the data ingestion pipeline [1].

Combining NiFi, Kafka, and Spark Streaming provides a compelling open source architecture option for building a next generation near real time ETL data pipelines. Next, we will evaluate the framework in a case study that involve ingesting and distributing global news articles from several sources for media monitoring.

## IV. EXPERIMENTAL EVALUATION

This section demonstrates how to consume massive data streams from streaming APIs using a global news and social media monitoring use case scenario.

*A. Use Case Scenario*

To illustrate the utility of the dataflow management framework, we use a media monitoring use case with the following functional requirements.
- Ingest streaming news articles data from several sources such as RSS feeds and Twitter Streaming API.
- Integrate the ingested streams with other news article sources being scraped continuously from a variety of specialized sources.
- Extract and filter noise such as fake news and duplicates in the data streams.
- Route news articles to relevant consumers such as persistent data stores or analytics engines.
- Support data enrichment, provenance, extensibility, and guaranteed message delivery.

In this case study, we'll demonstrate how to utilize the dataflow management framework in Fig. 2 to fetch, extract, enrich, integrate, and distribute live news stories from Twitter Streaming API and Satori channels (Big RSS and Worldwide Live Data) [16]. Next, we will describe how to achieve these tasks using NiFi-Kafka dataflow in a cluster.

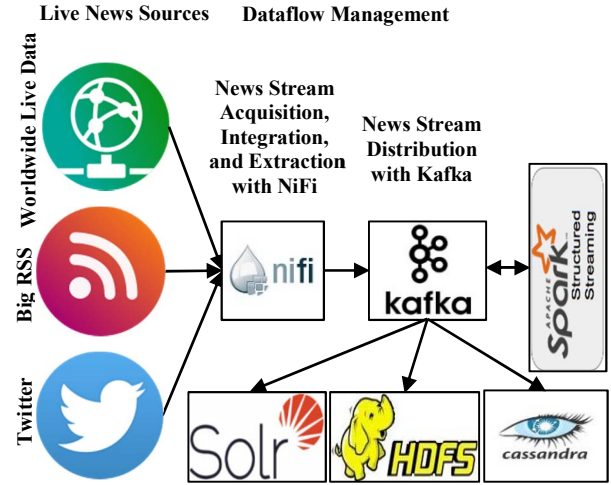

Figure 2. News articles processing infrastructure with a robust and scalable dataflow management framework

*B. Global News Articles Dataflow*

The main source of our data, Satori [16], is a cloud-based live platform, which provides a publish-subscribe messaging service called RTM and makes available a set of free real-time data feeds as part of their Open Data Channels initiative. We ingest news stories from Big RSS, a live data channel in Satori for gathering RSS feed. It is one of the largest RSS aggregators in the world with over 6.5 million feeds[3]. Another very important source of streaming news stories utilized is the Twitter API platform [17]. Twitter Inc offers a unified platform with scalable access to its data. There are currently two options (each with varying number of filters and filtering capabilities) for streaming real-time Tweets. We are used the standard/free option which allows 400 keywords, 5,000 user ids and 25 location boxes. There is also an enterprise option with premium operators that allows up to 250,000 filters (up to 2,048 characters each) per stream. The volume and velocity of data streams from the Twitter Streaming API depends on the popularity of the keyword queries. The entire flow was implemented using three local process groups in NiFi. Next, we describe the output of the dataflow framework and possible performance improvements.

*C. Dataflow Output and Performance Improvement*

We present the output of integrated news article sources (from Twitter, Big RSS, and custom WebSocket) that is continuously consumed and saved in HDFS. Fig. 3 shows a screenshot of recently processed news articles. NiFi automatically records, indexes, and makes available provenance data as objects through the system. Data can be downloaded, replayed, tracked and evaluated at numerous points along the dataflow path using the provenance user

---

[3] https://www.satori.com/livedata/channels/big-rss

interface. This information is useful for troubleshooting, optimization, and other scenarios.

![Figure 3 screenshot table]

Figure 3. Screenshot of processed news articles in HDFS

NiFi also provides a feature called data lineage (see Fig. 4), a visual representation of FlowFile content (generated from the news processing case study) which helps to track data from its origin to its destination.

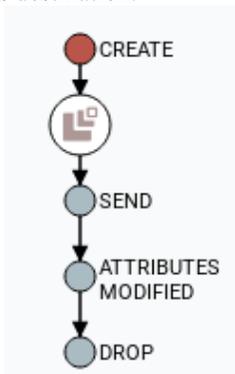

Figure 4. Data Lineage in NiFi

NiFi also provides a feature for graphical representation of the FlowFiles status history. This is useful in viewing various statics such as the number of bytes read, written, in, and out in 5 minutes. It also provides a feature called Back Pressure for indicating how much data should be allowed to exist in a queue before the source component is no longer scheduled to run. NiFi provides two configuration elements (object threshold and data size threshold) for Back Pressure. An object threshold specifies the maximum number of FlowFiles that should be queued up before applying back pressure (default value is 10,000 objects). A data size threshold specifies the maximum amount of data that should be queued up before applying back pressure (default value is 1 GB). Fig. 5 illustrates this concept when Kafka was down during our evaluation due to system maintenance and shows a maximum of 10,000 objects and red color.

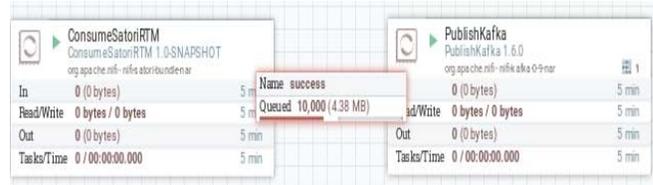

Figure 5. Back Pressure in NiFi

NiFi has several repositories, the database repository keeps track of all changes made to the dataflow using the user interface (UI). The FlowFile repository holds all the attributes of a FlowFile that are being processed. It allows NiFi to pick up where it left off in the event of a restart of the application due to unexpected power outage, inadvertent user/software restart, or upgrade. If FlowFile repository becomes corrupt or runs out of disk space, state of the FlowFiles can be lost. The content repository is where all the actual content of the files being processed by NiFi resides. It is also where multiple historical versions of the content are held if data archiving is enabled. Content repository can be very I/O intensive depending on the nature of the dataflow. The provenance repository keeps track of the lineage on both current and past FlowFiles. Like any application, the overall performance is governed by the performance of the individual components.

The basic NiFi configuration is far from ideal for high volume and high-performance dataflows. Some NiFi processors can be CPU, I/O and/or memory intensive. This is because NiFi is a Java application, and therefore, runs inside a Java Virtual Machine (JVM). Massive and high velocity flows of continuous data streams can cause performance bottlenecks in the memory, input/output (I/O) disk/network usage, or the central processing unit (CPU). However, the default NiFi core setting can be configured to run with improved performance.

V. RELATED WORK

Grover and Carey [7] developed an ingestion support (data feed) for a wide variety of data sources and applications in AsterixDB, an open-source big data management system. The system was designed to be used in AsterixDB and cannot be utilized for other use cases and applications outside AsterixDB. Park and Chi [18] proposed an open source based system for ingesting logs to a centralized HDFS clustered servers. The study lays a good foundation in this area, but the framework lacks in fundamental theoretical aspects and justification of the use of various components of the system. Meehan et al [11] discussed the major functional requirements of a streaming ETL. Their study, however, does not consider decoupling the ingestion system from the other components, which is highly recommended in modern and large-scale architectures.

Qiao et al [19] developed Gobblin, a generic data ingestion framework at LinkedIn. Gobblin was mainly driven by the fact that LinkedIn's data sources have become increasingly heterogeneous. It provides adaptors for commonly accessed data sources such as MySQL, S3, Kafka, and Salesforce. Other similar systems include Scribe [20], a messaging system as Facebook; Siphon [21], a messaging system in Microsoft

Azure HDInsight that utilize Kafka. Marcu et al [12] developed KerA, a data ingestion framework that alleviate the limitations of Kafka and other ingestion systems. The study focused on improving throughput, latency and scalability but do not consider issues such as provenance and extensibility.

Our study extends all these related works by utilizing open-source NiFi-Kafka integration to provide robustness and scalability as well as the ability to add and remove consumers at any time without changing the data ingestion pipeline.

## VI. CONCLUSIONS

Data ingestion is an essential part of companies and organizations that collect and analyze large volumes of data. Continuous data streams usually arrive into big data processing and management systems from external sources and are either incrementally processed or used to populate a persisted dataset and associated indexes. To keep pace with massive and fast-moving data, stream processing systems must be able to ingest, process, and persist data on a continuous basis. Developing an infrastructure for ingesting large-scale, multi-source, high velocity, and heterogeneous data streams involve a careful study of how these streams of data are produced and consumed.

A data stream ingestion system should be scalable, robust, and extensible to be able to support data flow between many independent data producers and consumers. This study proposes and developed a scalable and fault-tolerant dataflow management framework that can serve as a reusable component across many feeds of structured and unstructured input data. We demonstrated the utility of the framework in a real-world data stream processing case study that integrates Kafka and HDFS in a data flow system powered by NiFi. We showed sample outputs from our initial experimental work and discussed how the system can be configured for improved performance.

Future work will explore the deployment, evaluation, and practicality of improving the performance of the system by trying various NiFi and Kafka settings for different applications. We will also conduct comparative experiments with other open source and commercial data ingestion tools in order to provide a scientific study that can help the community in choosing the right framework for different applications.

ACKNOWLEDGMENT

Special thanks to Southern Ontario Smart Computing for Innovation Platform (SOSCIP) and IBM Canada for supporting this research project.